\documentclass[conference]{IEEEtran}
\IEEEoverridecommandlockouts
\usepackage{cite}
\usepackage{amsmath,amssymb,amsfonts}
\usepackage{algorithmic}
\usepackage{graphicx}
\usepackage{textcomp}
\usepackage{multirow}
\usepackage{xcolor}

\def\BibTeX{{\rm B\kern-.05em{\sc i\kern-.025em b}\kern-.08em
    T\kern-.1667em\lower.7ex\hbox{E}\kern-.125emX}}
\begin{document}

\title{CWGAN-GP Augmented CAE for Jamming Detection in 5G-NR in Non-IID Datasets

}
\author{\IEEEauthorblockN{Samhita Kuili, Mohammadreza Amini, Burak Kantarci} 
\IEEEauthorblockA{\textit{School of Electrical and Computer Engineering} \\
\textit{University of Ottawa}\\
Ottawa, Canada \\
Emails: \{skuil016, mamini6, burak.kantarci\}@uottawa.ca}
}

\maketitle

\begin{abstract}
In the ever-expanding domain of 5G-NR wireless cellular networks, over-the-air jamming attacks are prevalent as security attacks, compromising the quality of the received signal. We simulate a jamming environment by incorporating additive white Gaussian noise (AWGN) into the real-world In-phase and Quadrature (I/Q) OFDM datasets. A Convolutional Autoencoder (CAE) is exploited to implement a jamming detection over various characteristics such as heterogenous I/Q datasets; extracting relevant information on Synchronization Signal Blocks (SSBs), and fewer SSB observations with notable class imbalance. Given the characteristics of datasets, balanced datasets are acquired by employing a Conv1D conditional Wasserstein Generative Adversarial Network-Gradient Penalty(CWGAN-GP) 
on both majority and minority SSB observations. Additionally, we compare the performance and detection ability of the proposed CAE model on augmented datasets with benchmark models: Convolutional Denoising Autoencoder (CDAE) and Convolutional Sparse Autoencoder (CSAE). Despite the complexity of data heterogeneity involved across all datasets, CAE depicts the robustness in detection performance of jammed signal by achieving average values of 97.33\% precision, 91.33\% recall, 94.08\% F1-score, and 94.35 \% accuracy over CDAE and CSAE.
\end{abstract}

\vspace{-4mm}
\section{Introduction}
In recent years, 5G-NR wireless communication has been booming with a significant increase in wireless devices, for instance, smartphones, tablets, IoT, and massive IoT devices. With the advent of telecommunication infrastructure, wireless technologies encompass massive multiple input multiple output (MIMO)\cite{bjornson2020scalable}, millimeter-wave (mmwave)\cite{shen2019miniaturized}, carrier aggregation\cite{goyal2020lte}, learning-based resource allocation\cite{yu2020deep} 
which provision for end-to-end service connectivity between a 5G cellular network and end-users. On the contrary, a 5G-NR wireless cellular network is also susceptible to security attacks, 
notably jamming 
attacks, which 
intentionally 
disrupt signal-to-noise ratio, and bit error rate of the transmitted signals, degrading the communication quality. 
Jamming attacks target physical layer downlink channels and  downlink signals of 5G NR, 
exploiting the inherent vulnerabilities in Synchronization Signal Blocks (SSBs), which contain vital components like Primary and Secondary Synchronization Signals (PSS and SSS) responsible for cell identification and user association with gNodeB (gNB)\cite{giordani2018tutorial}. 


\begin{figure}[t]
\centerline{\includegraphics[scale = 0.6]{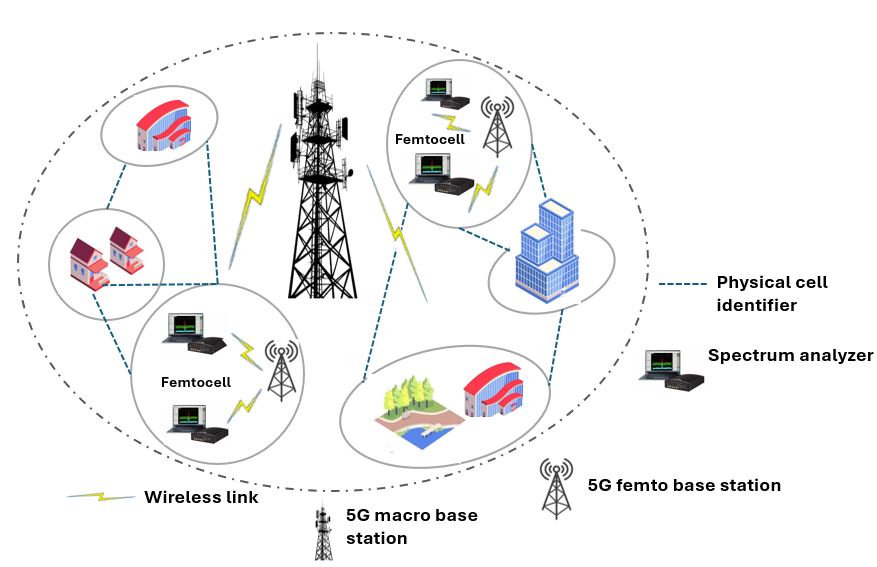}}
\caption{Jamming detection in a 5G-NR cellular network within a femtocell}
\label{fig1}\vspace{-6mm}
\end{figure}



A critical problem in 5G-NR networks is the heterogeneous data distribution from diverse user devices, as data is typically non-independent and identical distributed (non-IID) due to diverse geographical location. 
This causes the user datasets to vary significantly in size and data distribution across multiple users \cite{sattler2019robust}. As the 5G-NR network expands, jamming detection necessitates machine learning techniques \cite{Lohan.2024,hachimi2020multi} and deep learning on physical layer (PHY) to comprehend the underlying patterns of a propagated received signal. Existing deep learning-based detection methods assume uniform data distribution, which may not perfectly align with a real world 5G deployments where non-independent and identical distributed data is prevalent. We propose a jamming detection paradigm that takes into account for heterogeneous data obtained from each user while addressing class imbalance issues in real-world datasets. Varotto et al. \cite{varotto2023detecting} trains a convolutional autoencoder (CAE) only on non-jammed signals and proposes security strategies against attacks in orthogonal frequency-division multiplexing (OFDM)-based 5G signals. 
Additional models, such as the double-threshold deep neural network (DT-DDNN) \cite{Asemian2024DTDDNNAP} 
enable detection of wider types of jammers 
with lower false positive and miss detection rates by transforming I/Q samples into 2D images. Almazrouei et al. \cite{almazrouei2019using} propose a data-driven deep learning approach to denoise radio signals of IEEE 802.11 protocols without relying on expert knowledge by using convolutional denoising autoencoder and highlight an improvement in classification accuracy by exploiting both decoder and classifier. Luo et al. \cite{luo2017convolutional} propose a convolutional sparse autoencoder to sparsify the feature maps by integrating max-pooling into efficient feature leaning. These learned features are further used to propose a image classification strategy using the CSAE by integrating convolutional neural network. 

Jamming signals are rare, leading to significant class imbalance that results in poor deep learning performance in classifying non-jammed and jammed SSB signals. Varying channel conditions and interference levels alleviate the learning process. To address this challenge, our framework comprises Conditional Wasserstein Generative Adversarial Network with Gradient Penalty (CWGAN-GP) for augmenting minority class observations and mitigate data imbalance while CAE enhances feature extractions to improve classification performance. 
 Usage of Generative Adversarial Network (GAN) has been promising in effectively generating synthetic observations that closely resemble with the real data distribution and elevating the number of observation in the data. Chapaneri and Shah \cite{chapaneri2022enhanced, arjovsky2017wasserstein} discuss a reliable technique to enable data augmentation by exploiting a variant of GAN: Wasserstein GAN (WGAN) to improve the minority attack classification problem caused by cyber-attacks in network traffic. Chen et al.\cite{Chen2024ApplicationOG} 
use conditional Wasserstein generative adversarial network with gradient penalty (CWGAN-GP) based data augmentation to detect winding deformation in power transformers, and shows promising improvements over conventional Artificial intelligence (AI)-based fault diagnosis models. 
 A visual representation of femtocells in 5G-NR cellular network is shown in Fig. \ref{fig1}. The main contributions of the paper are highlighted below:

\begin{enumerate}
    \item A two-stage jamming detector tailored for 5G networks in RF domain is implemented by capturing  
     In-phase and Quadrature (I/Q) samples collected from over-the-air real-world 5G signals across multiple locations. 
    \item Unlike prior works which deal with uniform distribution and balanced datasets, we adopt CWGAN-GP to augment limited SSB observations focusing on non-IID datasets to mitigate the concern of class imbalance and ensuring more representative training distribution.
    \item The augmented datasets are further trained with CAE, which jointly executes both reconstruction and classification-based jamming detection, improving detection ability while addressing data heterogeneity across femtocells.
\end{enumerate}

Our work advances the existing state-of-the-art methods by adopting proposed framework and assessing the performance over benchmark models \cite{almazrouei2019using, luo2017convolutional} in identifying jammed signals while training on both non-jammed and jammed signals of a time domain dataset. The organization of this paper is as follows. 
Section \ref{M} elaborates on CWGAN-GP data augmentation technique for jamming detection. Section \ref{JD} discusses about the system model adopted for jamming detection. Section \ref{ES} presents the experimental setup with simulation results in Section \ref{ER}, and Section \ref{sec:con} summarizes the work in this article.  

\section{CWGAN-GP Augmented-based jamming detection }\label{M}
The objective of this work is to define an augmented ML-based approach which takes into account the dataset heterogeneity for each dataset collected at different geographical locations. This heterogeneity is identified by presence of non-IID data representing the attribute skewness, difference in quantity of SSB observations (training samples) across datasets, and imbalanced class distribution of jammed and non-jammed signals. The proposed framework deals with the stages of data collection and preprocessing to simulate a jammed 5G RF environment.
\subsection{Data collection}
Data is obtained with the help of spectrum analyzer which collects received signal waveform over-the-air, shared between telecommunication operators: Telus Communication Inc. and Rogers Communication Inc. Additionally, these received waveforms are acquired by setting a specific center carrier frequency 
and bandwidth 
over the available transmission cellular networks advocating various 5G-NR bands and bandwidths, respectively.

\subsection{Data preprocessing}
The collected received signal is transformed into spectrogram which coherently reflects the useful information of channel resource blocks. Only specific SSBs from resource blocks is extracted in the form of complex I/Q samples. Given \(\mathcal{N}\) different geographical locations, \(\mathcal{N}\) I/Q datasets are generated, each containing diverse training SSB observations. We assume the absolute values for I/Q samples which is effective for power-based jamming detection, where phase of the signal is ignored in the computation.  
Moreover, these absolute values are normalized across all datasets keeping a high-dimensional feature space. Furthermore, the incorporation of AWGN as jammed signal is simulated by varying the signal-to-noise (SNR) ratio to a suitable range for all the datasets. This provides information on the training SSBs with imbalanced class distribution of non-jammed and jammed signals across all datasets. Our proposed framework is not limited to AWGN but can also be leveraged for other types of jamming signals. 

\subsection{Data augmentation to tackle the class imbalance}
To tackle the data augmentation technique, a CWGAN-GP is chosen to generate more SSB observations as an oversampling approach. However, the oversampling is employed on both minority (non-jammed) and majority (jammed) signals to obtain a balanced binary classification problem. Additionally, augmentation facilitates CAE from becoming biased towards one class of signals. GAN consists of two neural networks (generator and discriminator) as proposed by Goodfellow et al. \cite{goodfellow2014generative}. The generator aims to leverage a Gaussian noise to obtain synthetic observations which resemble to the real data distribution. The objective function of a GAN follows a min-max game as formulated as, \vspace{-4mm}

\begin{equation}\label{eq:1}
\begin{split}
\min_G \max_D \textit{V(D, G)} = \hspace{0.1cm} &\mathbb{E}_{x_\sim p_{data}(x)}(log(\textit{D}(\textit{x}))\hspace{0.1cm} \\ 
& \> + \mathbb{E}_{z_\sim p_{z}(z)}(log(1-\textit{D}(\textit{G}(\textit{z})))
\end{split}
\end{equation}

The generator \textit{L$_G$} and the discriminator \textit{L$_D$} losses are represented as follows: \vspace{-8mm}

\begin{equation}\label{eq:2}
\begin{split}
\textit{L$_G$} = - \mathbb{E}_{z_\sim p_{z}(z)}(\textit{D}(\textit{G}(\textit{z}))))
\end{split}
\end{equation}

\vspace{-7mm}

\begin{equation}\label{eq:3}
\begin{split}
\textit{L$_D$} &= -\bigg[\mathbb{E}_{x_\sim p_{data}(x)}\Big(log\textit{D}(\textit{x})\Big)\hspace{0.1cm}  \\ & \hspace{10mm} + \mathbb{E}_{z_\sim p_{z}(z)}\Big(log(1-\textit{D}(\textit{G}(\textit{z}))\Big) \bigg]
\end{split}
\end{equation}

 where \textit{p$_{data}(x)$} denotes the real data distribution; \textit{p$_{z}(z)$} represents Gaussian distribution noise z; \textit{G($\cdot$)} represents the generator function; E($\cdot$) represents the expected function; \textit{D($\cdot$)} represents the discriminator function. The computation of \textit{L$_D$} takes into account both real and generated data while distinguishing between them as in (\ref{eq:3}). WGAN and WGAN-GP leverage a metric \textit{Earth-Mover} (EM) distance as the measure of the distance between real data distribution and generated data distribution which is better than \textit{Jensen-Shanon} (JS) divergence followed in conventional GANs. WGAN is highly effective in circumventing the issue of mode collapse. The EM distance is expressed as,\vspace{-4mm}

\begin{equation}\label{eq:4}
\begin{split}
W(\mathbb{P}_r, \mathbb{P}_g) = \inf_{\gamma \in \Pi(\mathbb{P}_r, \mathbb{P}_g)} \mathbb{E}_{(x,y) \sim \gamma} \left[ \|x - y\| \right]
\end{split}
\end{equation}

where $\Pi(\mathbb{P}_r, \mathbb{P}_g)$ denotes the entire joint probability distribution $\gamma(x, y)$ of real distribution $\mathbb{P}_r$, and generated data distribution $\mathbb{P}_g$. Moreover, $W(\mathbb{P}_r, \mathbb{P}_g)$ depicts the minimum cost required to transfer the mass while converting the distribution $\mathbb{P}_r$ into $\mathbb{P}_g$. Furthermore, EM distance is relatively useful in obtaining meaningful gradients for gradient descent training. 
The objective function between the generator (\textit{G}) and the critic (\textit{C}) (known as the discriminator) for WGAN is defined as, \vspace{-4mm}

\begin{equation}\label{eq:5}
\begin{split}
\min_G \max_D V(D, G) &= \mathbb{E}_{\mathbf{x} \sim \mathbb{P}_r}[\log D(\mathbf{x})] \\ 
& \> - \mathbb{E}_{\mathbf{x} \sim \mathbb{P}_g}[\log(1 - D(\mathbf{x}))]
\end{split}
\end{equation}

On the contrary, WGAN still fails to converge due to the weight clipping factor in WGAN. Therefore, Gulrajani \cite{gulrajani2017improved} introduces WGAN-GP, an extension of WGAN which penalizes the norm of the gradient of the critic concerning its input. This enables WGAN-GP to be appropriate for stable training with almost no hyperparameter tuning. The modified objective function of WGAN-GP is defined as, \vspace{-4mm}

\begin{equation}\label{eq:6}
\begin{split}
\min_G \max_D V(D, G) &= \mathbb{E}_{\mathbf{x} \sim \mathbb{P}_r}[D(\mathbf{x})] - \mathbb{E}_{\hat{\mathbf{x}} \sim \mathbb{P}_g}[D(\hat{\mathbf{x}})] \\ 
& \>   - \lambda \mathbb{E}_{\hat{\mathbf{x}} \sim \mathbb{P}_{\hat{\mathbf{x}}}} \left[ \left(\|\nabla_{\hat{\mathbf{x}}} D(\hat{\mathbf{x}})\|_2 - 1\right)^2 \right]
\end{split}
\end{equation}

where $\lambda$ is the gradient penalty coefficient $\hat{\mathbf{x}}$ is the sampling distributions between real distribution $\mathbb{P}_r$ and generated distribution $\mathbb{P}_g$ shown in (\ref{eq:7}):  \vspace{-4mm}

\begin{equation}\label{eq:7}
\begin{split}
\hat{\mathbf{x}} = \epsilon \mathbf{x} + (1 - \epsilon) \tilde{\mathbf{x}}, \quad \epsilon \sim \text{Uniform}[0, 1], \quad \mathbf{x} \sim \mathbb{P}_r, \quad \tilde{\mathbf{x}} \sim \mathbb{P}_g
\end{split}
\end{equation}

On the contrary, CWGAN-GP ensures auxiliary conditioned information $\mathbf{y}$; class label to both the critic and the generator. Formally, the objective value function that minimizes the loss function for the critic and the generator is expressed in (\ref{eq:8}), (\ref{eq:9}) and (\ref{eq:10}). \vspace{-4mm}

\begin{equation}\label{eq:8}
\begin{split}
\min_G \max_D V(D, G) = \mathbb{E}_{\mathbf{x} \sim \mathbb{P}_r} [D(\mathbf{x}|\mathbf{y})] - \mathbb{E}_{\tilde{\mathbf{x}} \sim \mathbb{P}_g} [D(\tilde{\mathbf{x}}|\mathbf{y})] 
 \\ - \lambda \mathbb{E}_{\hat{\mathbf{x}} \sim \mathbb{P}_{\hat{\mathbf{x}}}} \left[\left(\|\nabla_{\hat{\mathbf{x}}} D(\hat{\mathbf{x}}|\mathbf{y})\|_2 - 1\right)^2\right]
\end{split}
\end{equation}

\vspace{-4mm}

\begin{equation}\label{eq:9}
\begin{split}
L(D) = -\mathbb{E}_{\mathbf{x} \sim \mathbb{P}_r} [D(\mathbf{x}|\mathbf{y})] + \mathbb{E}_{\tilde{\mathbf{x}} \sim \mathbb{P}_g} [D(\tilde{\mathbf{x}}|\mathbf{y})]  \\ + \lambda \mathbb{E}_{\hat{\mathbf{x}} \sim \mathbb{P}_{\hat{\mathbf{x}}}} \left[\left(\|\nabla_{\hat{\mathbf{x}}} D(\hat{\mathbf{x}}|\mathbf{y})\|_2 - 1\right)^2\right]
\end{split}
\end{equation}
\vspace{-8mm}

\begin{equation}\label{eq:10}
\begin{split}
L(G) = -\mathbb{E}_{\tilde{\mathbf{x}} \sim \mathbb{P}_g} [D(\tilde{\mathbf{x}}|\mathbf{y})]
\end{split}
\end{equation}

\begin{figure}[t]
    \centering \includegraphics[width=.7\linewidth]{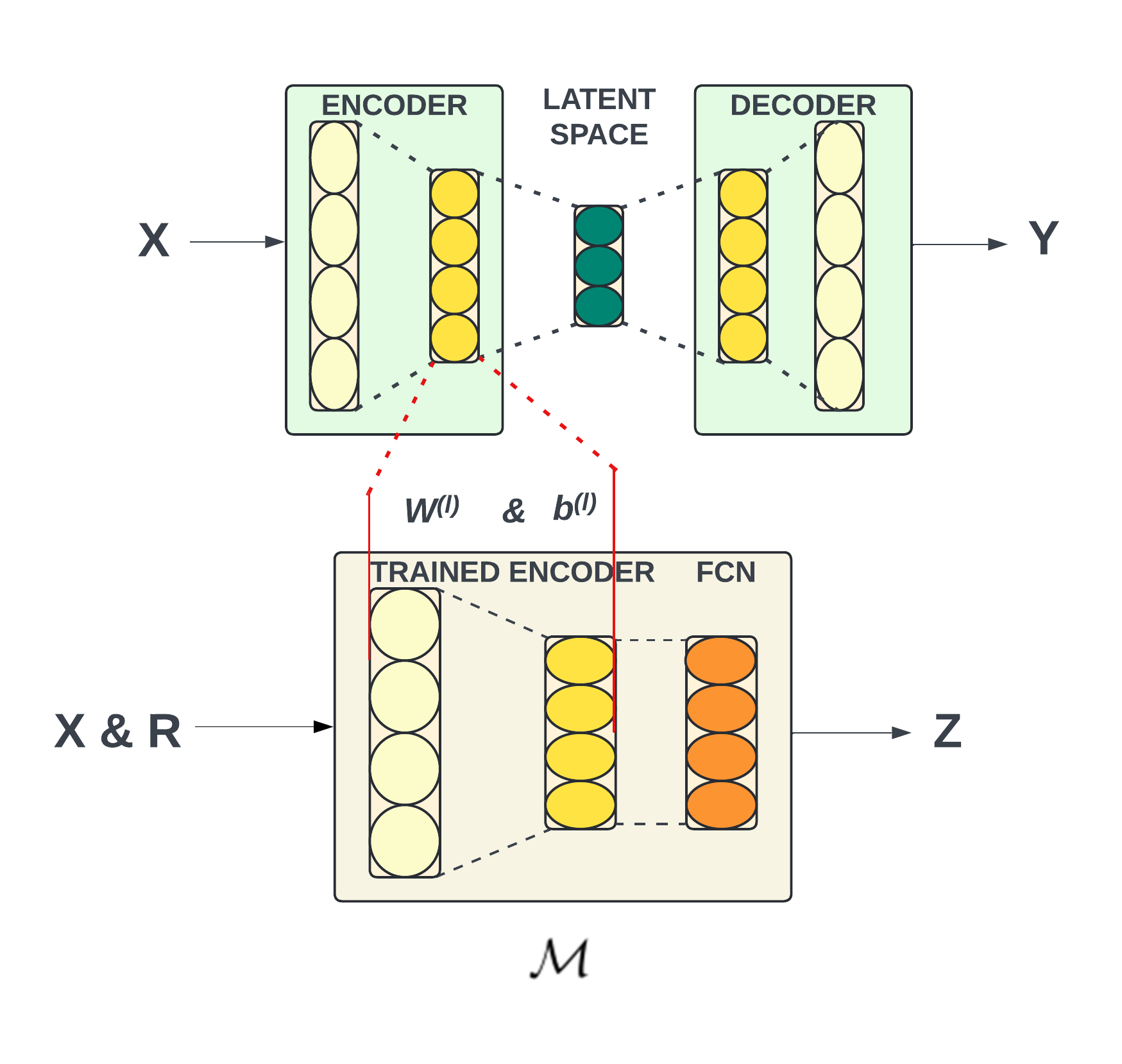}
    \caption{Architecture of convolutional autoencoder (CAE).}
    \label{CAE}
    \vspace{-2mm}
 \end{figure}

\section{Jamming detection with convolutional autoencoder}\label{JD}

The CAE is employed for one-class classification or jamming detection. The 2D temporal correlation in the augmented dataset is useful for undergoing a convolutional operation of the high-dimensional I/Q samples. Moreover, unlike other autoencoders where CAE is usually trained as a reconstruction, we intend to use CAE as both a reconstructor and a classifier. As illustrated in Fig. \ref{CAE}, CAE takes input array $\mathbb{\textbf{X}}$ of dimension {$\mathbb{\textbf{P}}$ by $\mathbb{\textbf{Q}}$; where $\mathbb{\textbf{P}}$ being SSB observations and $\mathbb{\textbf{Q}}$ is high-dimensional I/Q samples. The output for CAE is $\mathbb{\textbf{Y}}$, which is the same size as $\mathbb{\textbf{X}}$ due to the reconstruction characteristic of the model. The CAE comprises \textbf{\textit{L}} layers $\ell$ = 1,..., \textbf{\textit{L}}. The output of the final layer of encoder is obtained as (\ref{enc}). The decoder comprises transpose Conv1D layers, which form the reconstructed input from the encoded representation through compressed latent space. The output of the final layer of decoder is obtained as (\ref{dec}). \vspace{-4mm}

\begin{equation}\label{enc}
    \mathbf{U}^{(\ell)} = f\left(\mathbf{C}^{(\ell)} * \mathbf{U}^{(\ell-1)} + \mathbf{b}^{(\ell)}\right)
\end{equation}
\vspace{-8mm}

\begin{equation}\label{dec}
    \mathbf{V}^{(\ell)} = f\left(\mathbf{D}^{(\ell)} * \mathbf{V}^{(\ell-1)} + \mathbf{d}^{(\ell)}\right)
\end{equation}

where $\mathbf{U}^{(\ell)}$ and $\mathbf{V}^{(\ell)}$ are the outputs of the $\ell^{\text{th}}$ layer of encoder and decoder respectively, $f(\cdot)$ is the non-linear activation function, typically ReLU in this case. $\mathbf{C}^{(\ell)}$ and $\mathbf{D}^{(\ell)}$ are the convolutional weights at layer $\ell$, convolutional operation $\ast$ with $\mathbf{U}^{(\ell-1)}$ and $\mathbf{V}^{(\ell-1)}$, and $\mathbf{b}^{(\ell)}$, $\mathbf{d}^{(\ell)}$ as bias at layer $\ell$. The input of the first layer is $\mathbb{\textbf{X}}$ $\in$ $\mathbb{R}^{\textbf{P} \times \textbf{Q}}$ ,and the output of the last layer \textit{\textbf{L}} is $\mathbb{\textbf{Y}}$ = $\mathbf{V}^{(\textit{\textbf{L}})}$.

To implement jamming detection, our CAE is trained by compressing the input $\mathbb{\textbf{X}}$, representing the I/Q features of both jammed and non-jammed signals, using latent representation. The goal is to train the model in unsupervised learning to minimize the mean-square error (MSE) between $\mathbb{\textbf{X}}$ and $\mathbb{\textbf{Y}}$ as obtained in (\ref{loss}). However, the reconstructed weights $\textit{\textbf{W}}_{\text{e}}^{(\ell)}$ and  biases $\textit{\textbf{b}}_{\text{e}}^{(\ell)}$ from the trained encoder of CAE are captured from the $\ell^{\text{th}}$ layer of encoder. These weights and biases are transferred to the fully connected neural network (FCN); transforming the CAE to act as a classifier by combining trained encoder and FCN (added to the head of the encoder) into a new updated model $\mathcal{M}$ as shown in (\ref{weights}) and (\ref{biases}) respectively. \vspace{-3.5mm}

\begin{equation}\label{loss}
    \bar{\Gamma} = \mathbb{E}[\Gamma], \quad \Gamma = \|\mathbf{X} - \mathbf{Y}\|^2.
\end{equation}\vspace{-9mm}

\begin{equation}\label{weights}
    \textit{\textbf{W}}_{\mathcal{M}}^{(\ell)} = \textit{\textbf{W}}_{\text{e}}^{(\ell)}, \quad \forall \ell \in \{1, 2, \dots, L\}.
\end{equation}
\vspace{-9mm}

\begin{equation}\label{biases}
    \textit{\textbf{b}}_{\mathcal{M}}^{(\ell)} = \textit{\textbf{b}}_{\text{e}}^{(\ell)}, \quad \forall \ell \in \{1, 2, \dots, L\}.
\end{equation}

The detection ability of $\mathcal{M}$ is ensured by taking input $\mathbb{\textbf{X}}$ and ground truth $\mathbb{\textbf{R}}$, train it over 80\% train data and evaluate on 20\% test data with a suitable threshold $\gamma$.


\section{Experimental Setup}\label{ES}

An experimental setup is implemented within the 5G n71 band. As Per 3GPP specifications, this band spans a downlink frequency range from 617 MHz to 652 MHz, offering a total bandwidth of 35 MHz \cite{3gpp.38.104}. The frequency range is divided between two operators, TELUS and Rogers, each allocated 10 MHz of bandwidth. TELUS operates with a center frequency of 632 MHz, while Rogers operates at 622 MHz. The setup, depicted in Fig. \ref{fig_setup}, features a ThinkRF RTSA R5500 spectrum analyzer serving as the receiver with two different antennas to capture Over-The-Air (OTA) 5G signal from TELUS network. 

Sampling occurs at a frequency of 15.36 MHz across various environments, including indoor locations 
and outdoor scenarios (encompassing both Line-of-Sight (LOS) and Non-Line-of-Sight (NLOS) conditions). The gathered samples are saved in CSV format using the PyRF4 API and are subsequently processed. To obtain accurate information from the SSB, it is essential to estimate both the time offset (TO) and carrier frequency offset (CFO). Since the exact center frequency is unknown, a blind search approach is required. To precisely determine the TO and CFO, we leverage the PSS correlation properties and the cyclic prefix from the Cyclic Prefix Orthogonal Frequency Division Multiplexing (CP-OFDM) 5G waveform to align with the gNB signal.

\begin{figure}[t]
    \centering \includegraphics[width=.7\linewidth]{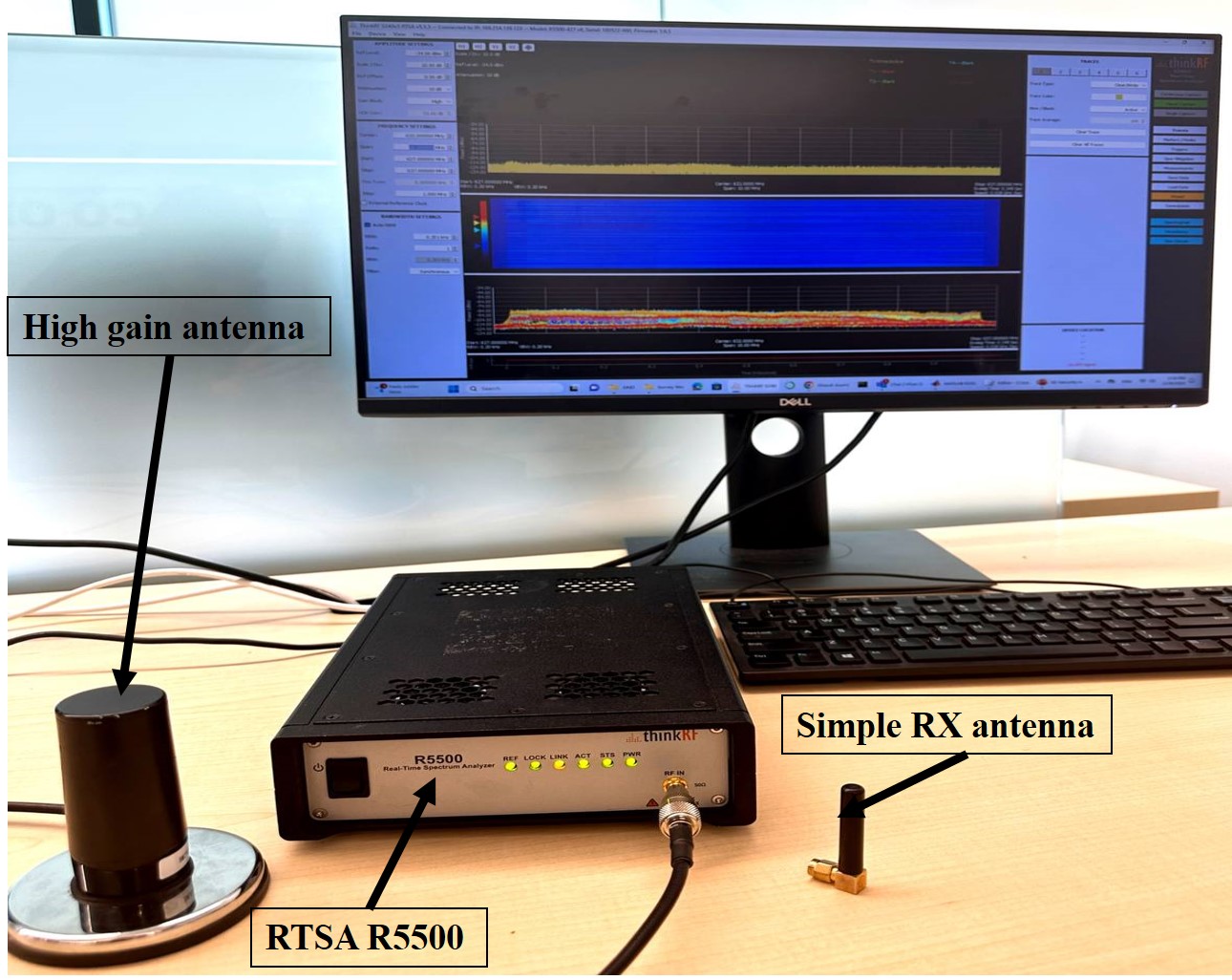}
    \caption{Experimental set-up for jamming detection.}
    \label{fig_setup}
    \vspace{-4mm}
 \end{figure}

The optimization problem for estimating the CFO is expressed as, \vspace{-9mm}

\begin{equation}\label{eq: optimization CFO}
\begin{split}
      \hat{\omega}_{CFO} &= \arg\max_{\omega_i} \Bigg[\sum_{\tau} y(\tau)e^{j \frac{\omega_i}{f_s}\tau}x_{pss}(t-\tau)   \Bigg],
\end{split}
\end{equation} where $x_{pss}$ is the primary synchronization signal, the first OFDM symbol in SSB and $f_s$ is the sampling frequency.
For obtaining time offset to the SSB, Schmidl \& Cox approach \cite{Schmidl1997} is used. Hence, the following optimization problem (\ref{eq: optimization TO}) is solved numerically where $\mathcal{P}(t)$ and $\mathcal{R}(t)$ are represented as (\ref{eq: TO-P}) and (\ref{eq: TO-R}), where $\hat{L}$ is one-half of the number of samples in one OFDM symbol. \vspace{-4mm}

\begin{equation}\label{eq: optimization TO}
\begin{split}
      \hat{T}_{off} &= \arg\max_{t} \, M(t)=\frac{|\mathcal{P}(t)|^2}{\mathcal{R}(t)^2} ,
\end{split}
\end{equation}

\vspace{-8mm}

\begin{equation}\label{eq: TO-P}
\begin{split}
      \mathcal{P}(t)=\sum_{n=0}^{\hat{L}-1} y^*(t+n)y(t+n+\hat{L})  ,
\end{split}
\end{equation}
 \vspace{-9mm}

\begin{equation}\label{eq: TO-R}
\begin{split}
      \mathcal{R}(t)=\sum_{n=0}^{\hat{L}-1}|y(t+n+\hat{L})|^2 
\end{split}
\end{equation}
\vspace{-5mm}

\begin{table}[t]
    \centering
    \caption{Information on datasets }\label{tab1}
    \resizebox{8cm}{!}{
    \begin{tabular}{|c|c|c|c|}
    \hline
             \normalsize{\textbf{Dataset ID}} &        \normalsize{\textbf{Location \& propagation conditions}} &  \normalsize{\textbf{SSB observation count}} & 
             \normalsize{\textbf{Class Imbalance}}\\ \hline
        \normalsize{1} &  \normalsize{Banchory} (\normalsize{Outdoor, NLOS, LOS}) &  \normalsize{826} &  \normalsize{(1) : 793 (0) : 33}\\ 
         \normalsize{2} &  \normalsize{Legget} (\normalsize{Outdoor, LOS}) &  \normalsize{544} &  \normalsize{(1) : 518 (0) : 26} \\ 
         \normalsize{3} &  \normalsize{Indoor\_2} (\normalsize{Indoor, LOS}) &  \normalsize{971} &  \normalsize{(1) : 933 (0) : 32} \\ 
         \normalsize{4} &  \normalsize{Indoor\_3} (\normalsize{Indoor, NLOS}) &  \normalsize{1038} &  \normalsize{(1) : 998 (0) : 40} \\ 
         \normalsize{5} &  \normalsize{Indoor\_4} (\normalsize{Indoor, NLOS}) &  \normalsize{877} &  \normalsize{(1) : 839 (0) : 38} \\ 
         \normalsize{6} &  \normalsize{Indoor\_5} (\normalsize{Indoor, NLOS}) &  \normalsize{989} &  \normalsize{(1) : 945 (0) : 44} \\ 
         \normalsize{7} &  \normalsize{Neighbor\_2} (\normalsize{Outdoor, LOS, NLOS}) &  \normalsize{805} &  \normalsize{(1) : 771 (0) : 34}\\ 
         \normalsize{8} &  \normalsize{Neighbor\_3} (\normalsize{Outdoor, NLOS}) &  \normalsize{923} &  \normalsize{(1) : 886 (0) : 37}\\ 
         \normalsize{9} &  \normalsize{Neighbor\_1} (\normalsize{Outdoor, LOS}) &  \normalsize{749} &  \normalsize{(1) : 719 (0) : 30}\\ 
         \normalsize{10} &  \normalsize{Park Shirley} (\normalsize{Outdoor, LOS, NLOS}) &  \normalsize{833} &  \normalsize{(1) : 799 (0) : 34} \\ 
         \normalsize{11} &  \normalsize{Shirin Market} (\normalsize{Outdoor, LOS})&  \normalsize{664} &  \normalsize{(1) : 638 (0) : 27} \\ 
         \normalsize{12} &  \normalsize{Stop Sign} (\normalsize{Outdoor, LOS}) &  \normalsize{978} &  \normalsize{(1) : 937 (0) : 41} \\ \hline
    \end{tabular}    
    }\ \vspace{-4mm}
\end{table}

\begin{table}[b]
    \vspace{-7mm}
    \centering
    \caption{CWGAN-GP Parameters and Hyperparameters}\label{tab2}
    \resizebox{6cm}{!}{
    \begin{tabular}{|c|c|}
    \hline
            \normalsize{\textbf{Parameter/Hyperparameter}} &  \normalsize{\textbf{Value/Details}} \\ \hline
            \normalsize{Model Architecture} &  \normalsize{\textit{C}: 32-512 units, \textit{G}: 128-64 units} \\ 
            \normalsize{Latent Vector Dimension} & \normalsize{128} \\ 
            \normalsize{Dropout} & \normalsize{\textit{C}: 0.5, \textit{G}: None} \\ 
            \normalsize{Batch Normalization}  & \normalsize{\textit{C}: None, \textit{G}: Yes} \\ 
            \normalsize{Activation Functions}  & \normalsize{\textit{C} and \textit{G} hidden: LeakyReLU, \textit{G} output: tanH} \\ 
            \normalsize{Batch Size}   & \normalsize{64} \\ 
            \normalsize{Training Epochs} &  \normalsize{20} \\
            \normalsize{Optimizer} & \normalsize{$\alpha$: 0.0001, $\beta_1$: 0.5, $\beta_2$: 0.9} \\
            \normalsize{Gradient Penalty Coefficient} & \normalsize{20} \\ 
            \normalsize{Critic Training} & \normalsize{7} \\ \hline
    \end{tabular}
    }\ \vspace{-2mm}
\end{table}

\section{Experimental Results}\label{ER}

The simulation is performed on 12 heterogeneous datasets, each comprising fewer SSB observations with a significant class imbalance of jammed (1) and non-jammed (0) signals. The information on each dataset is summarized in Table \ref{tab1}.

\subsection{Data augmentation using CWGAN-GP}

\begin{table*}[ht]
\centering
\caption{Parameters and hyperparameters of autoencoders}
\resizebox{11cm}{!}{
\begin{tabular}{|l|c|c|c|}
\hline
 \normalsize{\textbf{Parameter/Hyperparameter}} & \multicolumn{3}{c|}{ \normalsize{\textbf{Value/Details}}} \\ \hline
                                  &  \normalsize{\textbf{CAE}} &  \normalsize{\textbf{CDAE}} &  \normalsize{\textbf{CSAE}} \\ \hline
 \normalsize{Number of Layers (Encoder)}        &  \normalsize{3}            &  \normalsize{3}             &  \normalsize{3}             \\ 
 \normalsize{Number of Layers (Decoder)}        &  \normalsize{3}            &  \normalsize{3}             &  \normalsize{3}             \\ 
 \normalsize{Sparsity probability}              &  \normalsize{-}            &  \normalsize{-}             &  \normalsize{0.05}          \\ 
 \normalsize{Sparsity factor}                   &  \normalsize{-}            &  \normalsize{-}             &  \normalsize{0.01}          \\ 
 \normalsize{Noise factor}                      &  \normalsize{-}            &  \normalsize{0.3}           &  \normalsize{-}             \\ 
 \normalsize{Activation}                        &  \normalsize{ReLU}         &  \normalsize{ReLU}          &  \normalsize{ReLU}          \\ 
 \normalsize{Dropout}                           &  \normalsize{0.2}          &  \normalsize{0.2}           &  \normalsize{0.2}           \\ 
 \normalsize{Batch size}                        &  \normalsize{200}          &  \normalsize{200}           &  \normalsize{200}           \\ 
 \normalsize{Learning rate}                     &  \normalsize{0.0001}       &  \normalsize{0.0001}        &  \normalsize{0.0001}        \\ 
 \normalsize{Epochs}                            &  \normalsize{30 (Autoencoder \& Classifier)} &  \normalsize{15 (Autoencoder), 30 (Classifier)} &  \normalsize{15 (Autoencoder), 30 (Classifier)} \\ 
 \normalsize{Optimizer}                         &  \normalsize{Adam (Autoencoder \& Classifier)} &  \normalsize{Adagrad (Autoencoder), Adam (Classifier)} &  \normalsize{SGD (Autoencoder), Adam (Classifier)} \\ 
 \normalsize{Loss function}                     &  \normalsize{MSE and BCE}  &  \normalsize{MSE and BCE}   &  \normalsize{MSE and BCE}   \\ \hline
\end{tabular}\label{tab3}
}\ \vspace{-4mm}
\end{table*}

\begin{table*}[ht]
\centering
\caption{Jamming detection outcome comparison on 80:20 Training set/Testing set}
\resizebox{11cm}{!}{
\begin{tabular}{|c|ccc|c|c|ccc|c|c|ccc|c|c|}
\hline
\multirow{2}{*}{ \normalsize{Dataset ID}} & \multicolumn{5}{c|}{ \normalsize{CAE}} & \multicolumn{5}{c|}{ \normalsize{CDAE}} & \multicolumn{5}{c|}{ \normalsize{CSAE}} \\ \cline{2-16} 
                           &  \normalsize{Precision} & \normalsize{Recall} & \normalsize{F1-Score} & \normalsize{FAR} & \normalsize{MDR} & \normalsize{Precision} & \normalsize{Recall} & \normalsize{F1-Score} & \normalsize{FAR} & \normalsize{MDR} & \normalsize{Precision} & \normalsize{Recall} & \normalsize{F1-Score} & \normalsize{FAR} & \normalsize{MDR}\\ \hline
\normalsize{1}                          & \normalsize{100}       & \normalsize{82}     & \normalsize{90}       &\normalsize{0}     &\normalsize{17.8}       & \normalsize{83}        & \normalsize{98}     & \normalsize{90}      &\normalsize{19.9}    &\normalsize{2}     & \normalsize{97}        & \normalsize{95}     & \normalsize{96}    &\normalsize{2.7}    &\normalsize{5}       \\ 
\normalsize{2}                          & \normalsize{97}        & \normalsize{92}     & \normalsize{95}     &\normalsize{2.5}     &\normalsize{8}      & \normalsize{64}        & \normalsize{88}     & \normalsize{74}      &\normalsize{47.6}     &\normalsize{12}    & \normalsize{88}        & \normalsize{98}     & \normalsize{93}    &\normalsize{12.5}    &\normalsize{2}     \\ 
\normalsize{3}                          & \normalsize{97}        & \normalsize{81}     & \normalsize{88}       &\normalsize{2.7}      &\normalsize{19} 
 & \normalsize{85}        & \normalsize{96}     & \normalsize{90}       &\normalsize{15.5}     &\normalsize{4}   & \normalsize{93}        & \normalsize{92}     & \normalsize{92}     &\normalsize{7}    &\normalsize{8}  \\ 
\normalsize{4}                          & \normalsize{97}        & \normalsize{95}     & \normalsize{96}     &\normalsize{3.1}      &\normalsize{5}  & \normalsize{91}        & \normalsize{97}     & \normalsize{94}         &\normalsize{10.6}     &\normalsize{3}   & \normalsize{93}        & \normalsize{89}     & \normalsize{91}      &\normalsize{7.2}    &\normalsize{11} \\ 
\normalsize{5}                          & \normalsize{100}       & \normalsize{99}     & \normalsize{99}     &\normalsize{0.4}      &\normalsize{1}      & \normalsize{84}        & \normalsize{98}     & \normalsize{91}         &\normalsize{18}     &\normalsize{2}  & \normalsize{94}        & \normalsize{97}     & \normalsize{96}     &\normalsize{6}    &\normalsize{3}  \\ 
\normalsize{6}                          & \normalsize{92}        & \normalsize{95}     & \normalsize{94}      &\normalsize{8.1}      &\normalsize{5}     & \normalsize{98} & \normalsize{82}     & \normalsize{90}    &\normalsize{1.8}     &\normalsize{18}       & \normalsize{87}        & \normalsize{88}     & \normalsize{87}     &\normalsize{14.1}    &\normalsize{12}  \\ 
\normalsize{7}                          & \normalsize{100}       & \normalsize{99}     & \normalsize{99}    &\normalsize{0.4}      &\normalsize{1}   & \normalsize{94}        & \normalsize{90}     & \normalsize{92}     &\normalsize{6.2}     &\normalsize{10}  & \normalsize{98}        & \normalsize{98}     & \normalsize{98}    &\normalsize{2}    &\normalsize{2}   \\ 
\normalsize{8}                          & \normalsize{99}        & \normalsize{92}     & \normalsize{95}    &\normalsize{1.1}      &\normalsize{8}   & \normalsize{97}        & \normalsize{84}     & \normalsize{90}     &\normalsize{2.7}     &\normalsize{16}    & \normalsize{90}        & \normalsize{94}     & \normalsize{92}     &\normalsize{9.9}    &\normalsize{6}  \\ 
\normalsize{9}                          & \normalsize{92}        & \normalsize{68}     & \normalsize{78}     &\normalsize{6.4}      &\normalsize{32}  & \normalsize{97}        & \normalsize{95}     & \normalsize{96}     &\normalsize{2.6}     &\normalsize{5}    & \normalsize{95}        & \normalsize{97}     & \normalsize{96}     &\normalsize{5.1}    &\normalsize{3}  \\ 
\normalsize{10}                         & \normalsize{98}        & \normalsize{97}     & \normalsize{98}    &\normalsize{1.6}      &\normalsize{3}    & \normalsize{99}        & \normalsize{86}     & \normalsize{92}     &\normalsize{1}     &\normalsize{14}    & \normalsize{51}        & \normalsize{65}     & \normalsize{57}    &\normalsize{68.6}    &\normalsize{35}   \\ 
\normalsize{11}                         & \normalsize{100}        & \normalsize{99}     & \normalsize{100}    &\normalsize{0.1}      &\normalsize{1}   & \normalsize{92}        & \normalsize{91}     & \normalsize{91}    &\normalsize{7.6}     &\normalsize{9}   & \normalsize{98}        & \normalsize{95}     & \normalsize{96}     &\normalsize{1.9}    &\normalsize{5}  \\ 
\normalsize{12}                         & \normalsize{96}        & \normalsize{97}     & \normalsize{97}     &\normalsize{4.3}      &\normalsize{3}  & \normalsize{92}        & \normalsize{96}     & \normalsize{94}     &\normalsize{9.5}     &\normalsize{4}    & \normalsize{95}        & \normalsize{93}     & \normalsize{94}    &\normalsize{5.21}    &\normalsize{7}  \\ \hline
\end{tabular}\label{tab4}
}\ \vspace{-4mm}
\end{table*}

\begin{figure}[t]
    \centering \includegraphics[width=.6\linewidth]{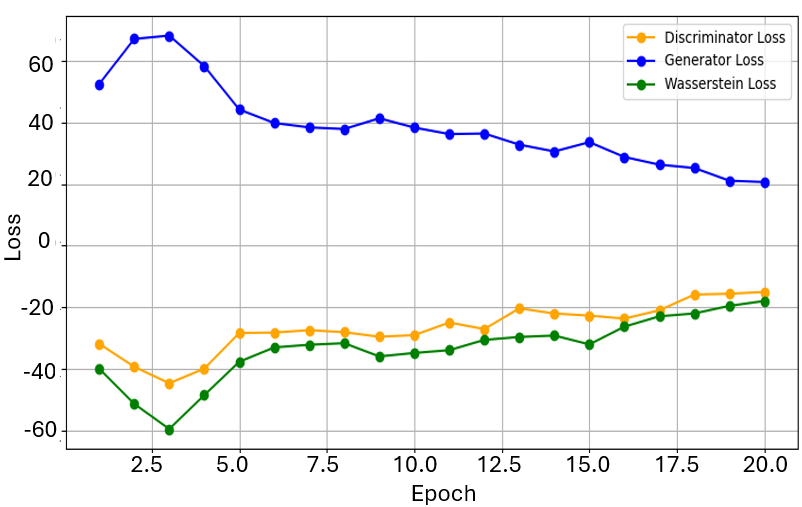}
    \caption{Training Loss Curves in CWGAN-GP.}
    \label{CWGAN_loss}
    \vspace{-4mm}
 \end{figure}

We adopt CWGAN-GP, which handles heterogeneity on each dataset by augmenting the number of observations to a fixed amount, for instance, 5000 observations; enforcing 2500 jammed and 2500 non-jammed signals. The entire class imbalance for each dataset is assumed to be the training set prior to oversampling using CWGAN-GP. The architecture of CWGAN-GP follows a five-layer Conv1D neural networks for \textit{C} and two Conv1D neural networks for \textit{G}. 
CWGAN-GP is trained over a few epochs with a fixed batch size \cite{mckeever2020synthesising} 
to generate 250 generated observations i.e. 5000 observations; which comprise 2500: jammed and 2500: non-jammed SSB observations. We choose the default values for optimizer Adam, set gradient penalty coefficient, and train critic a few times unlike the default values used in \cite{gulrajani2017improved}. Table \ref{tab2} presents the details on parameters and hyperparameters for CWGAN-GP. The CWGAN-GP model shows convergence over the training epochs (see Fig. \ref{CWGAN_loss}), depicting critic loss stabilizes along with Wasserstein loss. However, generator loss spikes during the early stages of training, highlighting that the generated samples are far from real samples, and gradually stabilizes over time to generate more realistic samples.   

\subsection{Training with CAE, CDAE and CSAE}
CAE is trained on each dataset ID to showcase detection performance in terms of classification metric precision, recall, F1-score, and accuracy of the model. However, jamming detection requires other metrics, for instance, False Alarm Rate  (FAR) and Missed Detection Rate (MDR) to comprehend the real-world deployment effectiveness. FAR and MDR metrics are critical for measuring false alarms and potential indications of compromising network security. Moreover, CAE is first trained in an unsupervised learning algorithm while assuming 8:2 as training and validation sets. During the first training process, CAE captures the weights and biases of the trained encoder and is transferred to the fully connected layer; acting as a classifier, and subsequently trained in a supervised learning manner. The parameters and hyperparameters for the CAE model are highlighted in Table \ref{tab3}. The jamming detection performance of the classifier using the trained weights showcases promising accuracy, precision, recall, and f1-score obtained for each dataset ID while considering $\gamma$ = 0.5. However, Dataset ID 9 achieves lower recall and F1-score of 68\% and 78\% respectively as compared to other datasets. This signifies that a larger proportion of true jammed signals are incorrectly detected as false negatives or non-jammed signals. In addition, the missed detection rate is 0.32 which depicts that 32\% of jammed signals are identified as non-jammed signals. Moreover, the false alarm rate is 0.064 or 6.4\% of true non-jammed signals are incorrectly identified as jammed signals.

On the contrary, CDAE \cite{almazrouei2019using} and CSAE \cite{luo2017convolutional} are trained unsupervised and compute reconstruction errors between the input samples and the decoded output. Only the reconstruction errors are used at the input to the trained encoder and fully connected layer to obtain the classification performance with the same threshold unlike the similar training followed for CAE. However, the weights/biases are captured by CDAE and CSAE and forwarded to FCN similar to CAE. The detection ability of CDAE shown in Table \ref{tab4} highlights promising performance across all datasets but Dataset ID 2; achieving a precision, recall, and F1-score of 64\%, 88\%, and 74\%, respectively. The low value of precision depicts the presence of high false positives. The lower false negative provides a direct hint of obtaining a higher recall. In addition, the missed detection rate for Dataset ID 2 shows that 12\% of jammed signals are identified as non-jammed signals and the false alarm rate of 47.6\% of non-jammed signals are incorrectly identified as jammed signals; causing more false positives. On the contrary, CSAE performs satisfactorily well across all the datasets but Dataset ID 10 with precision, recall, F1-score, and accuracy are shown in Table \ref{tab4}. The poor detection performance coherently indicates high false negatives and high false positives responsible for acquiring low precision and recall, respectively. In terms of missed detection rate and false alarm rate, 35\% of the jammed signals are distinguished as non-jammed signals, and 68.6\% of the non-jammed signals are mistaken as jammed signals. The performance differences across all datasets are evident due to varying propagation and channel conditions of jamming power at different locations.  In addition, the accuracy comparison for the models across all the datasets highlights CAE outperforms CDAE and CSAE shown in Fig. \ref{accuracy}. Moreover, a comparison showcases notable performance differences by assuming the proposed CAE over the other benchmark models: CDAE and CSAE. The average of precision, F1-score, and accuracy highlight that the proposed CAE model outperforms the benchmark models with a significant difference shown in Table \ref{tab5}. 

\begin{figure}[t]
    \centering \includegraphics[width=0.55\linewidth]{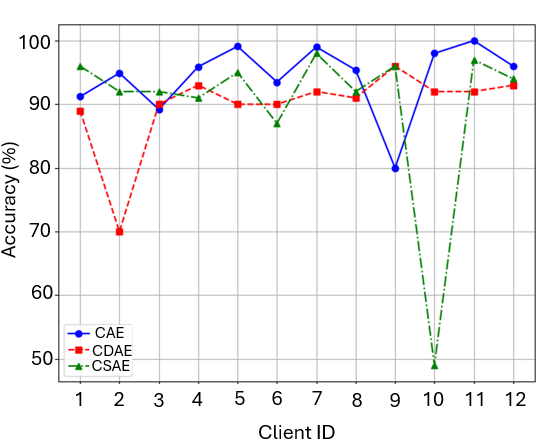}
    \caption{Accuracy comparison of each dataset.}
    \label{accuracy}
    \vspace{-4mm}
 \end{figure}

 \begin{table}[t]
    \centering
    \caption{Average classification performance metrics of models}
     \resizebox{6cm}{!}{
    \begin{tabular}{|c|c|c|c|c|}
    \hline
        \textbf{Models} & \textbf{Precision (\%)} & \textbf{Recall (\%)} & \textbf{F1-score (\%)} & \textbf{Accuracy (\%)} \\ \hline
        \textbf{CAE} & 97.33 & 91.33 & 94.08 & 94.35 \\ \hline
        \textbf{CDAE} & 89.67 & 91.75 & 90.33 & 89.93 \\ \hline
        \textbf{CSAE} & 89.92 & 91.75 & 90.67 & 89.92 \\ \hline
    \end{tabular}\label{tab5}
     }\vspace{-4mm}
    
\end{table}

\section{Conclusion and Future Work}\label{sec:con}
We have proposed an augmented-based jamming detection against 5G-NR networks while assuming various factors: data heterogeneity across multiple femtocells, limited SSB observations, and the presence of class imbalance across all datasets. Our approach employs the exploitation of CWGAN-GP to generate more synthetic SSB observations and obtain balanced datasets; comprising an equal amount of jammed and non-jammed signals. To ensure high classification performance and detection of jammed attacks, we employ CAE and train the model in both unsupervised and supervised learning on IQ signals of a 5G-NR cellular network. The results depict that the detection ability of CAE outperforms other benchmark models: CDAE and CSAE in terms of metrics: precisions, acceptable recall, F1-score, and accuracy. However, a detailed comparison of CAE model over benchmark models across all datasets showcases that the proposed approach performs better by achieving an accuracy of at least 90\% without the involvement of reconstruction errors in the training process unlike CDAE and CSAE. The detection performance of CAE relies on the quality of augmented samples of CWGAN-GP, which might impact the performance if there is frequent fluctuation of generator loss without converging over time. Our ongoing work aims to address computational complexity and optimization strategies 
to improve scalability by assuming more femtocells in a 5G-NR network. 

\section*{Acknowledgment}
This work was supported in part 
by the Natural Sciences and Engineering Research Council of Canada (NSERC) under the Discovery and CREATE TRAVERSAL Programs.

\bibliographystyle{IEEEtran}

\begin{thebibliography}{10}
\providecommand{\url}[1]{#1}
\csname url@samestyle\endcsname
\providecommand{\newblock}{\relax}
\providecommand{\bibinfo}[2]{#2}
\providecommand{\BIBentrySTDinterwordspacing}{\spaceskip=0pt\relax}
\providecommand{\BIBentryALTinterwordstretchfactor}{4}
\providecommand{\BIBentryALTinterwordspacing}{\spaceskip=\fontdimen2\font plus
\BIBentryALTinterwordstretchfactor\fontdimen3\font minus \fontdimen4\font\relax}
\providecommand{\BIBforeignlanguage}[2]{{%
\expandafter\ifx\csname l@#1\endcsname\relax
\typeout{** WARNING: IEEEtran.bst: No hyphenation pattern has been}%
\typeout{** loaded for the language `#1'. Using the pattern for}%
\typeout{** the default language instead.}%
\else
\language=\csname l@#1\endcsname
\fi
#2}}
\providecommand{\BIBdecl}{\relax}
\BIBdecl

\bibitem{bjornson2020scalable}
E.~Bj{\"o}rnson and L.~Sanguinetti, ``Scalable cell-free massive mimo systems,'' \emph{IEEE Transactions on Communications}, vol.~68, no.~7, pp. 4247--4261, 2020.

\bibitem{shen2019miniaturized}
X.~Shen, Y.~Liu, L.~Zhao, G.-L. Huang, X.~Shi, and Q.~Huang, ``A miniaturized microstrip antenna array at 5g millimeter-wave band,'' \emph{IEEE Antennas and Wireless Propagation Letters}, vol.~18, no.~8, pp. 1671--1675, 2019.

\bibitem{goyal2020lte}
A.~Goyal and K.~Kumar, ``Lte-advanced carrier aggregation for enhancement of bandwidth,'' in \emph{Advances in VLSI, Communication, and Signal Processing: Select Proceedings of VCAS 2018}.\hskip 1em plus 0.5em minus 0.4em\relax Springer, 2020, pp. 341--351.

\bibitem{yu2020deep}
P.~Yu, F.~Zhou, X.~Zhang, X.~Qiu, M.~Kadoch, and M.~Cheriet, ``Deep learning-based resource allocation for 5g broadband tv service,'' \emph{IEEE Transactions on Broadcasting}, vol.~66, no.~4, pp. 800--813, 2020.

\bibitem{giordani2018tutorial}
M.~Giordani, M.~Polese, A.~Roy, D.~Castor, and M.~Zorzi, ``A tutorial on beam management for 3gpp nr at mmwave frequencies,'' \emph{IEEE Communications Surveys \& Tutorials}, vol.~21, no.~1, pp. 173--196, 2018.

\bibitem{sattler2019robust}
F.~Sattler, S.~Wiedemann, K.-R. M{\"u}ller, and W.~Samek, ``Robust and communication-efficient federated learning from non-iid data,'' \emph{IEEE Trans on neural networks and learning systems}, vol. 31/9, pp. 3400--3413, 2019.

\bibitem{Lohan.2024}
P.~Lohan, B.~Kantarci, M.~Amine~Ferrag, N.~Tihanyi, and Y.~Shi, ``From 5g to 6g networks: A survey on ai-based jamming and interference detection and mitigation,'' \emph{IEEE Open Journal of the Communications Society}, vol.~5, pp. 3920--3974, 2024.

\bibitem{hachimi2020multi}
M.~Hachimi, G.~Kaddoum, G.~Gagnon, and P.~Illy, ``Multi-stage jamming attacks detection using deep learning combined with kernelized support vector machine in 5g cloud radio access networks,'' in \emph{Intl. Symp. on networks, computers and communications}.\hskip 1em plus 0.5em minus 0.4em\relax IEEE, 2020, pp. 1--5.

\bibitem{varotto2023detecting}
M.~Varotto, S.~Valentin, and S.~Tomasin, ``Detecting 5g signal jammers with autoencoders based on loose observations,'' in \emph{2023 IEEE Globecom Workshops (GC Wkshps)}.\hskip 1em plus 0.5em minus 0.4em\relax IEEE, 2023, pp. 160--165.

\bibitem{Asemian2024DTDDNNAP}
\BIBentryALTinterwordspacing
G.~Asemian, M.~Amini, B.~Kantarci, and M.~Erol-Kantarci, ``Dt-ddnn: A physical layer security attack detector in 5g rf domain for cavs,'' \emph{ArXiv}, vol. abs/2403.02645, 2024. [Online]. Available: \url{https://api.semanticscholar.org/CorpusID:268249163}
\BIBentrySTDinterwordspacing

\bibitem{almazrouei2019using}
E.~Almazrouei, G.~Gianini, C.~Mio, N.~Almoosa, and E.~Damiani, ``Using autoencoders for radio signal denoising,'' in \emph{Proceedings of the 15th ACM International Symposium on QoS and Security for Wireless and Mobile Networks}, 2019, pp. 11--17.

\bibitem{luo2017convolutional}
W.~Luo, J.~Li, J.~Yang, W.~Xu, and J.~Zhang, ``Convolutional sparse autoencoders for image classification,'' \emph{IEEE transactions on neural networks and learning systems}, vol.~29, no.~7, pp. 3289--3294, 2017.

\bibitem{chapaneri2022enhanced}
R.~Chapaneri and S.~Shah, ``Enhanced detection of imbalanced malicious network traffic with regularized generative adversarial networks,'' \emph{J. of Network and Computer Applications}, vol. 202, p. 103368, 2022.

\bibitem{arjovsky2017wasserstein}
M.~Arjovsky, S.~Chintala, and L.~Bottou, ``Wasserstein generative adversarial networks,'' in \emph{International conference on machine learning}.\hskip 1em plus 0.5em minus 0.4em\relax PMLR, 2017, pp. 214--223.

\bibitem{Chen2024ApplicationOG}
\BIBentryALTinterwordspacing
Y.~Chen, Z.~Zhao, J.~Liu, S.~Tan, and C.~Liu, ``Application of generative ai-based data augmentation technique in transformer winding deformation fault diagnosis,'' \emph{Engineering Failure Analysis}, 2024. [Online]. Available: \url{https://api.semanticscholar.org/CorpusID:267641919}
\BIBentrySTDinterwordspacing

\bibitem{goodfellow2014generative}
I.~Goodfellow, J.~Pouget-Abadie, M.~Mirza, B.~Xu, D.~Warde-Farley, S.~Ozair, A.~Courville, and Y.~Bengio, ``Generative adversarial nets,'' \emph{Advances in neural information processing systems}, vol.~27, 2014.

\bibitem{gulrajani2017improved}
I.~Gulrajani, F.~Ahmed, M.~Arjovsky, V.~Dumoulin, and A.~C. Courville, ``Improved training of wasserstein gans,'' \emph{Advances in neural information processing systems}, vol.~30, 2017.

\bibitem{3gpp.38.104}
3GPP, ``{5G; NR; Base Station (BS) radio transmission and reception},'' {3rd Generation Partnership Project (3GPP)}, Technical Specification (TS) 38.104, 02 2024, version 17.12.0.

\bibitem{Schmidl1997}
T.~Schmidl and D.~Cox, ``Robust frequency and timing synchronization for ofdm,'' \emph{IEEE Transactions on Communications}, vol.~45, no.~12, pp. 1613--1621, 1997.

\bibitem{mckeever2020synthesising}
S.~McKeever and M.~S. Walia, ``Synthesising tabular datasets using wasserstein conditional gans with gradient penalty (wcgan-gp),'' \emph{Technological University Dublin}, 2020.

\end{thebibliography}

\end{document}